# FEDSM2026

# N.E.O.N.-BRIDGE GEOMETRY DETERMINATION: TURBULENCE MODELING OF INDIVIDUAL N.E.O.N.-BRIDGE SEGMENT

**Arturo Rodriguez[1], Dominic Alexander[2], Nicolas J. Torres[1], Benay Ozcelik[1], Omar Escudero[2], Ty Reitzel[2], Pablo Rangel[2]**

[1]Texas A&M University – Kingsville, Kingsville, Texas, United States of America
[2]Texas A&M University – Corpus Christi, Corpus Christi, Texas, United States of America

**ABSTRACT**

The N.E.O.N.-Bridge is a capstone project being developed by students at Texas A&M University – Corpus Christi, under the oversight of Los Alamos National Laboratory. The project requires the development of a hull geometry for an autonomous bridge segment optimized to support onboard electronics and camera systems while maintaining stability in a dynamic water environment. Traditional ribbon bridge systems typically do not experience intense hydrodynamic loading due to current transportation and assembly methods, whereas the N.E.O.N-Bridge must continuously withstand forces from dynamic flow patterns. The requirement for a hull geometry to have both hydrodynamic design and rigidity, as in current ribbon bridges, has posed unique challenges.

The current hull designs were evaluated through turbulent water-flow simulations performed with ANSYS Discovery. Boundary conditions were determined based on the forward motion of the bridge segment, simulating inlet and outlet flows, and the resulting pressure distribution. A waterline-based geometric constraint derived from the camera system's elevation enabled the simulation to model flow characteristics in an operational scenario. Velocity fields, pressure contours, and turbulent flow patterns were analyzed to identify areas of high loading and hydrodynamic inefficiencies.

The findings will provide essential performance metrics that could be used to make design adjustments to the overall hull geometry. The simulation results will support improvements in stability, structural rigidity, and overall effectiveness of the hull geometry, advancing the development of the N.E.O.N-Bridge segments.

**NOMENCLATURE**

| | |
|---|---|
| $\rho$ | density |
| t | time |
| u | x-component of velocity |
| x | x-coordinate |
| $v$ | y-component of velocity |
| y | y-coordinate |
| w | z-component of velocity |
| z | z-coordinate |
| p | pressure |
| g | gravity |
| $\mu$ | dynamic viscosity |
| $v_t$ | eddy viscosity |
| $S_{ij}$ | shear stress transport matrix |
| k | turbulent kinetic energy |
| $\delta_{ij}$ | kronecker delta |
| $\omega$ | turbulent dissipation |
| $P_k$ | turbulent production |
| $\beta, \alpha, \sigma_k, \sigma_\omega$ | model constants |

## 1. INTRODUCTION

Floating bridge systems remain a critical asset for rapid military troop mobility, disaster response, and civil infrastructure applications. Conventional bridges and troop structures are typically designed under the assumption that hydrodynamic forces play a secondary role, with operational speed and control

 

of deployment conditions being the primary limitations. However, emerging concepts for autonomous bridges necessitate a fundamental re-examination of these assumptions. Modular bridge segments operating continuously in dynamic river environments must withstand sustained forces from river hydrodynamics, turbulent flow, and structural loads, all of which directly influence stability, segment maneuverability, and sensor performance [1, 2].

The N.E.O.N-Bridge (Neuromorphic Event-Driven Optical Networking Bridge) represents a paradigm shift in floating bridge design. Unlike traditional bridges, the N.E.O.N. bridge is envisioned as autonomous, capable of self-propulsion coordinated with its continuous movement, alignment, and operation in environments with GPS or RF-denied systems. This operational requirement introduces the problem of fluid-structure coupling, where the geometry must simultaneously satisfy hydrodynamic efficiency, structural rigidity, and the functional integration of electronics, cameras, and communication systems [3].

From a fluid mechanics perspective, the hydrodynamic behavior of a bridge segment is governed by turbulent flow, with interactions characterized by the Reynolds number at which our bridge operates. These interactions include boundary layer development on flat and curved surfaces, flow separation at abrupt geometric transitions, wake formation, and pressure recovery behind the segment. This results in a distribution of pressures and shear forces that directly affects the drag force, pitching moments, and lateral stability—critical qualities for maintaining alignment during autonomous operation [4]. Consequently, predicting turbulence using turbulence models is essential for guiding optimal geometry selection and evaluating performance.

Reynolds-Averaged Navier-Stokes (RANS) modeling remains the most practical and robust approach for turbulence analysis in engineering applications. Higher-fidelity modeling techniques, such as Large Eddy Simulations (LES) or Direct Numerical Simulations (DNS), offer a more physically accurate representation. Still, their computational cost remains prohibitive for design studies involving complex geometries and multiple operating conditions. RANS models, when correctly applied, provide an effective balance between physical accuracy and computational efficiency, enabling the systematic exploration of velocity fields, pressure distributions, and turbulence characteristics relevant to hull performance [5, 6, 7].

In this work, we focus on modeling the turbulence of an individual segment of the N.E.O.N bridge using RANS numerical simulations performed in ANSYS Discovery [8]. The analysis will be framed operationally, with realistic boundary conditions representing water movement; the inlet and outlet conditions capture the characteristics of the dominant flow experienced by the segment during deployment and handling in the river. A water-based geometric constraint is derived from the camera and elevation sensor placement and their requirements; it is incorporated into the numerical simulation fields to reflect the system's operational conditions. Velocity distributions, pressure contours, and turbulent flow structures are examined to identify regions affected by buoyancy, flow separation, and hydrodynamic inefficiencies [9].

The objective of this study is twofold. First, it aims to establish a physics-based analysis of the hydrodynamic performance of candidate geometries under turbulent flow conditions. Second, it seeks to inform the design process with insights that will guide geometry refinement, ensure stability, and reduce adverse loads without compromising the structural requirements and operational functionalities of autonomous floating bridges [10, 11]. The design will be modeled with turbulence in mind; this work contributes to integrating fluid mechanics, autonomy, and systems engineering into the next generation of floating bridge infrastructure [12, 13].

## 2. MATERIALS AND METHODS

### 2.1 Geometry

The geometry of the N.E.O.N. bridge was carefully designed to balance hydrodynamic performance, structural integrity, and the accommodation of systems within the bridge segment. Unlike conventional elements of traditional bridges, which are primarily optimized for static buoyancy and modular assembly, the N.E.O.N. bridge segment must operate continuously in a dynamic flow environment. As a result, geometry must account for the active interaction between the fluid and the structure, rather than simply floating on the water's surface. The selected geometries feature a combination of planar and curved surfaces, designed to predict boundary-layer development and minimize adverse pressure gradients that could lead to flow separation and time-dependent loads. Attention was paid to the hull taper profile in regions that strongly influence pressure stagnation, wake formation, and drag characteristics.

Geometric constraints can be imposed on system requirements, particularly regarding elevation and the requirements of the hardware sensing cameras in Figure 1 and Figure 2. These constraints define the hull geometry, which directly influences hull depth, freeboard, and wetted surface area. The result is a geometry that reflects the design philosophy, where hydrodynamic efficiency is pursued not only for its own sake, but also through the integration of sensor placement, internal volume requirements, and structural stiffness. Smooth geometric transitions reduce localized turbulence and pressure spikes, while flat regions provide the necessary support for internal components and facilitate manufacturing. The geometry provides a fundamental physical basis for turbulence modeling, enabling numerical simulations that capture relationships among shape, flow, and operational performance in an autonomous environment.

　

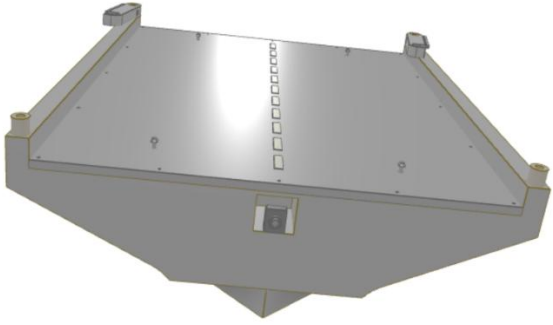

**Fig. 1** Plain N.E.O.N. Bridge Segment Geometry

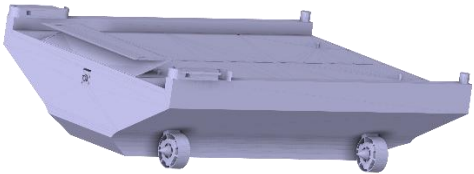

**Fig. 2** Actual N.E.O.N. Bridge Segment Geometry

### 2.2 Navier-Stokes Equations

The hydrodynamic behavior of the N.E.O.N. bridge segment is governed by the Navier-Stokes equations, which express the fundamental conservation laws of mass and momentum for a viscous, incompressible flow. These equations form the physical basis for describing the evolution of the velocity and pressure fields around the bridge geometry, which interacts with the surrounding water and its motion. In this application, the flow is characterized by high Reynolds numbers, dominated by viscous stresses that develop in the boundary layer and near the wall, and by its turbulent behavior. The incompressibility assumption is appropriate for underwater operation under the conditions considered; the density is constant and is decoupled from the pressure field.

In differential form, these equations governing the flow consist of the continuity equation, which represents the conservation of mass, and the momentum equations, which represent Newton's second law applied to the fluid elements. The nonlinear convective terms capture the momentum transport by the flow and the emergence of flow instabilities and turbulent regions, where accuracy can change drastically with geometric complexity. Viscous diffusion results from molecular viscosity and the critical interplay between shear stresses and the moving surface. Together, the equations describe the balance between pressure forces, inertial transport, and viscous dissipation that ultimately dictates the drag force, lift, and stability characteristics of the bridge segment.

When applied to realistic geometries, such as the N.E.O.N. bridge segment, the Navier-Stokes equations cannot be solved analytically and therefore must be solved numerically. The geometric complexity, combined with the nature of turbulence and flow, necessitates a numerical solution. The Navier-Stokes equations are used to derive turbulence models. This formulation ensures that numerical simulations are grounded in fundamental physical principles, providing meaningful interpretations of velocity fields, pressure distributions, and force-generating mechanisms relevant to the autonomous operation of floating bridge segments.

$$\frac{\partial \rho}{\partial t} + \rho \left( \frac{\partial u}{\partial x} + \frac{\partial v}{\partial y} + \frac{\partial w}{\partial z} \right) = 0 \qquad (1)$$

$$\rho \left( \frac{\partial u}{\partial t} + \frac{\partial u}{\partial x} u + \frac{\partial u}{\partial y} v + \frac{\partial u}{\partial z} w \right) \qquad (2)$$
$$= -\frac{\partial p}{\partial x} + \mu \left( \frac{\partial^2 u}{\partial x^2} + \frac{\partial^2 u}{\partial y^2} + \frac{\partial^2 u}{\partial z^2} \right) + \rho g_x$$

$$\rho \left( \frac{\partial v}{\partial t} + \frac{\partial v}{\partial x} u + \frac{\partial v}{\partial y} v + \frac{\partial v}{\partial z} w \right) \qquad (3)$$
$$= -\frac{\partial p}{\partial y} + \mu \left( \frac{\partial^2 v}{\partial x^2} + \frac{\partial^2 v}{\partial y^2} + \frac{\partial^2 v}{\partial z^2} \right) + \rho g_y$$

$$\rho \left( \frac{\partial w}{\partial t} + \frac{\partial w}{\partial x} u + \frac{\partial w}{\partial y} v + \frac{\partial w}{\partial z} w \right) \qquad (4)$$
$$= -\frac{\partial p}{\partial z} + \mu \left( \frac{\partial^2 w}{\partial x^2} + \frac{\partial^2 w}{\partial y^2} + \frac{\partial^2 w}{\partial z^2} \right) + \rho g_z$$

### 2.3 Reynolds Decomposition

The Reynolds decomposition provides a systematic framework for representing turbulent flow by separating the instantaneous flow into a mean and a fluctuating component, offering a time-averaged statistical description of the flow dynamics. In this procedure, the velocity and pressure fields are expressed as the sum of the mean flow and fluctuating components with zero expectation, reflecting the unsteady nature of turbulence while preserving its dominant and repeatable characteristics. Substituting this decomposition into the Navier-Stokes equations and subsequently averaging them yields equations governing the mean flow and the effects of turbulence, which appear as additional terms in the stress tensor and result in correlations among the velocity fluctuations. These Reynolds stresses result from turbulent eddy transport and play a central role in determining wave development, pressure recovery, and force distribution in the N.E.O.N. segment. The Reynolds decomposition introduces the closure problem, which establishes the physical and computational basis for turbulence modeling, particularly for the analysis of complex geometries operating at high Reynolds numbers.

$$\rho = \bar{\rho} + \rho' \tag{5}$$

$$\boldsymbol{u} = \bar{\boldsymbol{u}} + \boldsymbol{u}' \tag{6}$$

$$\overline{\rho'} = 0 \tag{7}$$

$$\overline{\boldsymbol{u}'} = 0 \tag{8}$$

### 2.4 Reynolds-Averaged Navier-Stokes (RANS) Turbulence Modeling

The Reynolds-averaged Navier-Stokes (RANS) equations provide a practical, physically grounded framework for predicting turbulent flows in engineering applications where resolving all turbulence scales is neither feasible nor necessary. By applying the Reynolds decomposition to the Navier-Stokes equations and time-averaging, the governing equations are expressed in terms of mean flow quantities, with the effects of turbulence represented by the Reynolds stress tensor. These additional stress terms encapsulate momentum transport due to fluctuations and are modeled using a closure scheme. At high Reynolds numbers, which characterize the operating environment of the N.E.O.N. bridge, the Reynolds stresses dominate over molecular viscosity and influence boundary layer behavior, wave formation, and pressure distributions during movement. RANS models offer a practical compromise between physical fidelity and computational efficiency, enabling systematic evaluation for complex geometries under realistic operating conditions while maintaining a direct connection to physical conservation principles. RANS models provide robust modeling of hydrodynamic forces, stability, and performance, which are essential for the design and refinement of the bridge segment's motion.

$$\frac{\partial \bar{\rho}}{\partial t} + \nabla \cdot (\bar{\rho}\bar{\boldsymbol{u}} + \overline{\rho'\boldsymbol{u}'}) = 0 \tag{9}$$

$$\frac{\partial[\bar{\rho}\bar{u} + \rho'u']}{\partial t} + \frac{\partial[(\bar{\rho})(\bar{u}^2 + u'^2)]}{\partial x} \tag{10}$$
$$+ \frac{\partial[(\bar{\rho})(\bar{u}\bar{v} + u'v')]}{\partial y}$$
$$+ \frac{\partial[(\bar{\rho})(\bar{u}\bar{w} + u'w')]}{\partial z}$$
$$= -\frac{\partial(\bar{p} + p')}{\partial x} + \mu\nabla^2(\bar{u} + u') + (\bar{\rho} + \rho')g_x$$

$$\frac{\partial[\bar{\rho}\bar{v} + \rho'v']}{\partial t} + \frac{\partial[(\bar{\rho})(\bar{v}^2 + v'^2)]}{\partial x} \tag{11}$$
$$+ \frac{\partial[(\bar{\rho})(\bar{v}\bar{u} + v'u')]}{\partial y}$$
$$+ \frac{\partial[(\bar{\rho})(\bar{v}\bar{w} + v'w')]}{\partial z}$$
$$= -\frac{\partial(\bar{p} + p')}{\partial y} + \mu\nabla^2(\bar{v} + v') + (\bar{\rho} + \rho')g_y$$

$$\frac{\partial[\bar{\rho}\bar{w} + \rho'w']}{\partial t} + \frac{\partial[(\bar{\rho})(\bar{w}^2 + w'^2)]}{\partial x} \tag{12}$$
$$+ \frac{\partial[(\bar{\rho})(\bar{w}\bar{u} + w'u')]}{\partial y}$$
$$+ \frac{\partial[(\bar{\rho})(\bar{w}\bar{v} + w'v')]}{\partial z}$$
$$= -\frac{\partial(\bar{p} + p')}{\partial z} + \mu\nabla^2(\bar{w} + w') + (\bar{\rho} + \rho')g_z$$

### 2.5 $k - \omega$ Turbulence Model

The k-omega turbulence model is used to close the Reynolds-averaged Navier-Stokes equations, with emphasis on resolving near-wall behavior in regions with adverse pressure gradients. This two-equation system introduces transport equations for turbulent kinetic energy, k, and the specific dissipation rate, and establishes a link between the local and dynamic response and the formulation of the turbulent eddy viscosity. Unlike models that use empirical wall functions, the k-omega formulation resolves the viscous sublayer directly, making it suitable for applications where boundary-layer development, separation, and reattachment are central. These characteristics are critical for capturing the hydrodynamic response of the N.E.O.N. bridge maneuvers, which involve transitions and flow curvature, with strong pressure effects and significant stability.

From a design and analysis perspective, the k-omega model provides a balance between physical fidelity and computational efficiency at high Reynolds numbers, typical of rivers. By explicitly accounting for turbulence production and dissipation as functions of the mean strain and vorticity, the model captures the dominant mechanisms governing wake formation, shear layer growth, and momentum redistribution around the bridge segment. This capability is significant for time-dependent drag forces, pitching moments, and weights, with implications for autonomous alignment and maneuverability. As a result, the k-omega model serves as an effective tool for evaluating candidate geometries and guiding design, maintaining a direct connection between the physics of turbulent flow and the flow-structure interaction relevant to the operation of autonomous floating systems.

$$\rho\left(\frac{\partial \bar{u}_i}{\partial t} + \bar{u}_j \frac{\partial \bar{u}_i}{\partial x_j}\right) = -\frac{\partial \bar{p}}{\partial x_i} + \mu \frac{\partial^2 \bar{u}_i}{\partial x_j \partial x_j} - \frac{\partial}{\partial x_j}(\rho \overline{u'_i u'_j}) \quad (13)$$

$$-\rho \overline{u'_i u'_j} = 2\rho v_t S_{ij} - \frac{2}{3}\rho k \delta_{ij} \quad (14)$$

$$S_{ij} = \frac{1}{2}\left(\frac{\partial \bar{u}_i}{\partial x_j} + \frac{\partial \bar{u}_j}{\partial x_i}\right) \quad (15)$$

$$k = \frac{1}{2}\overline{u'_i u'_i} \quad (16)$$

$$v_t \sim \frac{k}{\omega} \quad (17)$$

$$\omega \sim \frac{\varepsilon(dissipation)}{k(turbulent\ creation)} \quad (18)$$

$$\frac{\partial k}{\partial t} + \bar{u}_j \frac{\partial k}{\partial x_j} = P_k - \beta^* k\omega + \frac{\partial}{\partial x_j}\left[(v + \sigma_k v_t)\frac{\partial k}{\partial x_j}\right] \quad (19)$$

$$\frac{\partial \omega}{\partial t} + \bar{u}_j \frac{\partial \omega}{\partial x_j} = \alpha \frac{\omega}{k} P_k - \beta \omega^2 + \frac{\partial}{\partial x_j}\left[(v + \sigma_\omega v_t)\frac{\partial \omega}{\partial x_j}\right] \quad (20)$$

$$P_k = 2v_t S_{ij} S_{ij} \quad (21)$$

## 3. RESULTS AND DISCUSSION

Numerical simulations provide detailed information on the hydrodynamic response of the N.E.O.N. bridge segment under turbulent conditions, representing the segment in operation. The velocity fields reveal the development of the boundary layer on the surface, followed by flow separation in regions where the geometry transitions and induces adverse pressure gradients (refer to Figure 3). These features are near the edges and downstream of the flat sections, where an interaction between the curvature and the incoming flow produces local acceleration and deceleration zones. The wake structure behind the segment is characterized by reduced average velocity and elevated turbulence intensity, indicating a significant momentum deficit and enhanced mixing driven by large eddies. These flow characteristics are consistent with expectations for bluff and semi-streamlined bodies at high Reynolds numbers and highlight regions where geometric refinement is needed to reduce drag and time-dependent loads.

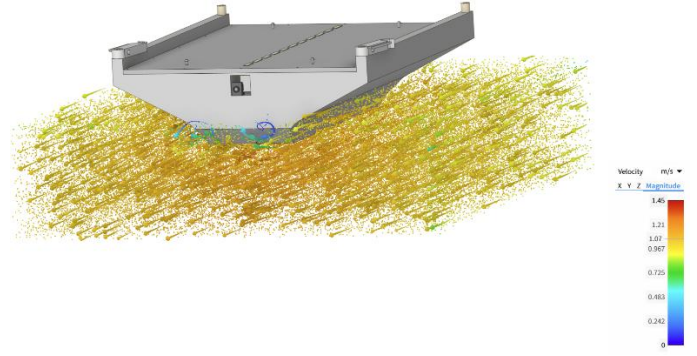

**Fig. 3** Isometric Velocity Vector Field

Figure 4 shows the velocity contours on one side of the N.E.O.N. bridge segment for the flow conditions indicated in the color map, revealing the primary characteristics of the velocity field in the vicinity of the movement. Below, a deceleration region is observed at the front of the segment, corresponding to the formation of the stagnation zone located beneath the moving surface. Further downstream from this region, the flow accelerates towards the inclined surface before the transition characterized by waves in the zone where the velocity magnitude decreases. The asymmetry of the velocity contours near the geometric transitions indicates the presence of adverse pressure gradients that promote an increase in boundary-layer thickness and premature separation, critical characteristics for determining hydrodynamic loads and drag forces.

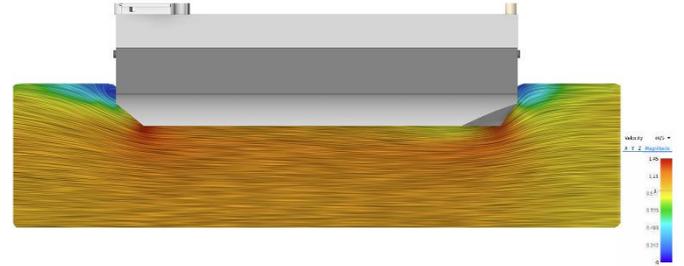

**Fig. 4** Side View Velocity Contour

In accordance with the vector field shown in the adjacent section of Figure 5, this provides additional information about the mechanisms governing the flow in momentum transport. The vectors highlight the redistribution of momentum beneath the hull, near the predominantly horizontal flow in the frontal region, and show the increasing inclination of the vectors towards the rear of the flow, where it conforms to the conical shape of the geometry. In some regions, they rise with a divergent vector aligned with low-velocity zones, as shown in Figure 4, indicating separation and high shear flow. These characteristics indicate the production of turbulence driven by sharp velocity gradients, reinforcing the need to resolve near-wall turbulence in this configuration using turbulence models.

 

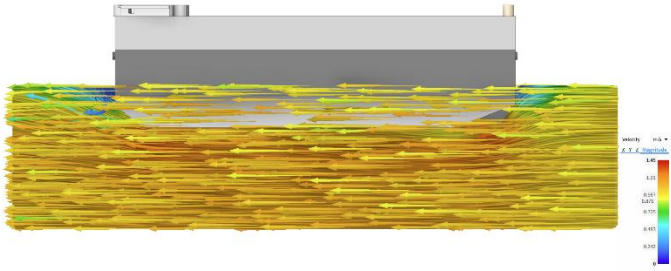

**Fig. 5** Side View Vector Field

The three-dimensional effects become more evident in the isometric visualizations shown in Figures 6 and 7. Figure 6 combines velocity vectors with magnitude contours to illustrate the interaction around the hull geometry and the flow surrounding it. These results show the deflection of the coherent flow around the inclined surfaces, accompanied by local acceleration near the sharper geometric features. In contrast, Figure 7 emphasizes the evolution of the flow, showing vector dispersion and reduced alignment, indicating the development of turbulent flow structures. The persistence of organized structures near the hull, followed by rapid downstream dissipation of the flow, suggests that geometry promotes partial flow attachment before the transition to turbulent flow. This behavior has direct implications for reducing drag and time-dependent forces.

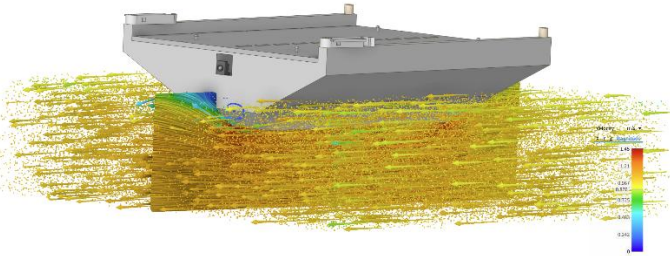

**Fig. 6** Isometric Velocity Vector Field with Contour

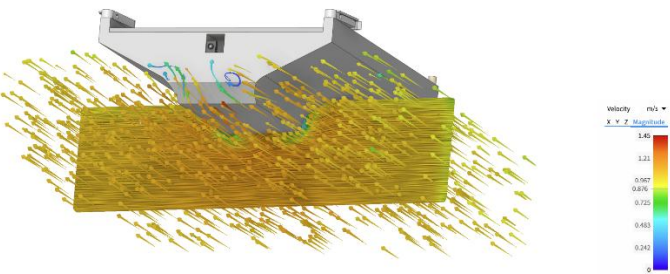

**Fig. 7** Isometric Velocity Vector Field with Contour II

Figure 8 presents a point cloud representation of the velocity field and its contours, offering a volumetric perspective of the turbulent flow near the bridge segment. The spatial distribution of the points reveals prominent regions of elevated velocity intensity, particularly beneath the deck and in the turbulent wake downstream. The concentration of points in these regions reflects increased turbulence activity and enhanced mixing, which is consistent with the distribution of Reynolds stresses inferred from the RANS solution. From a design perspective, these regions represent areas where time-dependent flow, along with its associated forces and vibrations, influences the structural response and sensor performance, underscoring the importance of refining geometry and mitigating adverse flow effects.

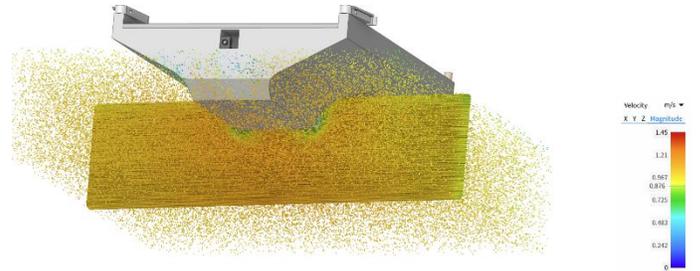

**Fig. 8** Point Cloud with Contour

Figure 9 represents a modified vector field illustrating the global redistribution of momentum resulting from the introduction of propulsion. In particular, the passive flow has been discussed previously; the velocity vectors show an increase behind the hull, indicating the partial suppression of the low-momentum regions previously associated with stagnation and separation. This redistribution suggests that the propulsion does not provide thrust directly, but instead alters the near-field hydrodynamics, thereby improving flow attachment and reducing wave asymmetry.

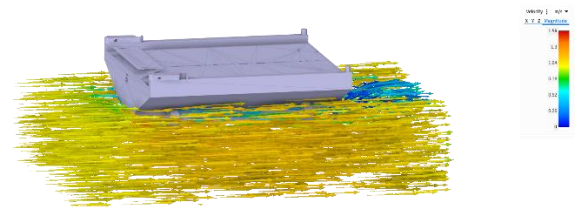

**Fig. 9** Modified Velocity Vector Field

The localized effects of the propulsion system are clearly revealed in Figure 10, in the velocity contours. Distinct regions of high velocity are observed in the vicinity of the propulsion systems, accompanied by vortex structures that indicate rotational flow and its vorticity. These propulsion jets induce vorticity in the surrounding flow, generating downstream vorticity patterns that interact with the vehicle. This symmetry in the velocity contours indicates the balance of propulsion forces, an essential requirement for maintaining directional stability during autonomous operation. At the same time, the intensity of the change in velocity in these regions highlights areas of high turbulence production and potential time-dependent forces.

 

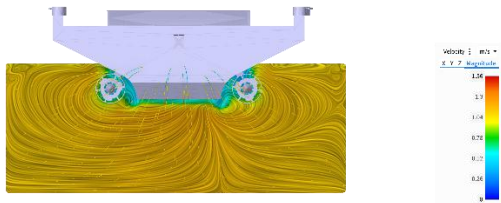

**Fig. 10** Back View with Propellers Velocity Contour

Figure 11 shows the isometric view of the velocity field in the mid-plane of the bridge segment, providing a three-dimensional perspective of the propulsion system's interaction with the vehicle hull. The contours reveal the accelerated flow induced by the propulsion systems near the hull, which significantly mitigates the low-recirculation zones identified in the cases without propulsion. Although the interaction between the propulsion jets and the hull geometry introduces localized regions of high shear, this particularly affects the characteristics and placement of the propulsion systems. These regions represent a trade-off between momentum transport and increased turbulence intensity, with implications for hydrodynamic efficiency and structural loads.

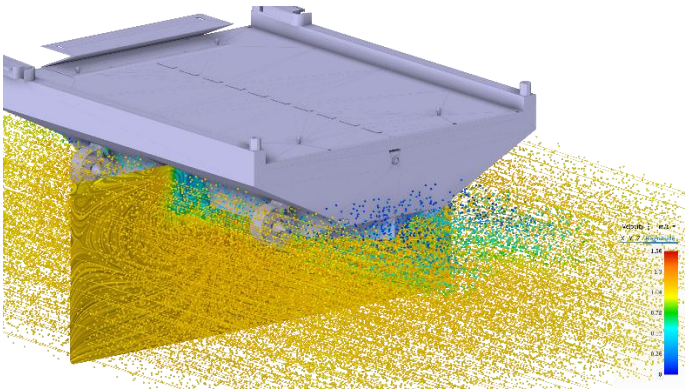

**Fig. 11** Middle Velocity Contour with Propellers

The side view of the velocity contours shown in Figure 12 clearly demonstrates the redistribution of vortices at a given moment in the propulsion systems. Compared to passive configurations, the presence of propulsion systems increases the velocity near the hull and promotes an energetic boundary-layer flow beneath the surface. This behavior suggests pressure recovery and wave reduction, both of which are favorable for drag reduction and vehicle control. At the same time, the contours indicate adverse pressure gradients that persist near the geometric transitions, highlighting the importance of the hull shape in actively manipulating the flow and controlling it through propulsion.

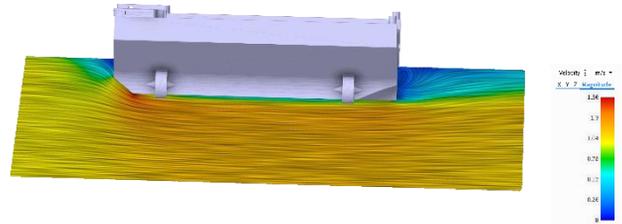

**Fig. 12** Side View Velocity Contour with Propellers

Figure 13 presents an isometric point cloud visualization with superimposed velocity contours, offering a volumetric perspective of the flow structures developing downstream of the N.E.O.N. bridge segment. This point distribution is notable for its spatial resolution of the waves. It reveals regions of high-velocity fluctuations immediately behind the hull, where the induced propulsion interacts with the waves generated by the geometry. The high-velocity areas near the edges indicate acceleration, driven by both propulsion and geometric effects, while the rapid downstream dispersion of the points reflects turbulent mixing and momentum diffusion. The antisymmetric observed in the velocity contours near the edges suggests the formation of a shear layer and the presence of coherent structures that contribute to the time-dependent forces and wave dispersion. From a physical standpoint, the visualization illustrates the three-dimensional nature of turbulence, highlighting the importance of volumetric wave effects when predicting the hydrodynamic performance, stability, and autonomous control of the bridge segment operating in turbulent environments.

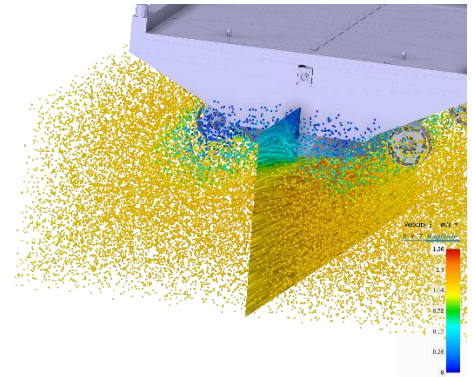

**Fig. 13** Back Isometric Viewpoint Cloud with Velocity Contour

## 4. CONCLUSION

This study demonstrates that Reynolds-averaged Navier-Stokes numerical simulations using the k-ω turbulence model achieve physical consistency and computational efficiency for evaluating the hydrodynamic performance of an individual N.E.O.N bridge segment operating in turbulent flow. The results establish a

  

connection between the hull geometry, turbulent flow structures, and hydrodynamic forces, revealing the behavior of the boundary layer, flow separation, and wave development, as well as the propulsion-induced moment for stability and drag force characteristics. This analysis demonstrates that active propulsion significantly modifies the flow field, mitigating momentum regions by introducing local shear force into the integrated design of the geometry and propulsion system. Collectively, these findings provide physical validation and an iterative foundation for refining the autonomous bridge segments and support the development of the next generation of floating bridge systems in dynamic river environments.

## ACKNOWLEDGEMENTS

Texas A&M University at Kingsville (TAMUK), Faculty Arturo Rodriguez Start-Up Funds. The U.S. Department of Energy (DOE) Grande CARES Consortium funded this research under grant number GRANT13584020.